\documentclass[pra, 10pt, twocolumn, tightenlines, longbibliography, aps, superscriptaddress,nofootinbib]{revtex4-2}

\usepackage{amsmath,amssymb,amsfonts,bm}
\usepackage{braket}
\usepackage[utf8]{inputenc}
\usepackage[T1]{fontenc}
\usepackage{float}
\usepackage[english]{babel}
\usepackage[normalem]{ulem}
\usepackage{mathtools}
\usepackage{bbold}
\usepackage{graphicx}
\usepackage{multirow}
\usepackage{tabularx}
\usepackage[dvipsnames, table]{xcolor}
\usepackage{soul}
\usepackage[colorlinks=true,allcolors=blue]{hyperref}
\usepackage{amsmath}
\newcommand{\abs}[1]{\left| #1 \right|} 

\begin{document}

\title{Correlated many-body quantum dynamics of the Peregrine soliton}

\author{D. Diplaris}
\affiliation{Center for Optical Quantum Technologies, Department of Physics, University of Hamburg, Luruper Chaussee 149, D-22761 Hamburg, Germany}
\affiliation{The Hamburg Centre for Ultrafast Imaging, University of Hamburg, Luruper Chaussee 149, D-22761 Hamburg, Germany}
 
\author{G. A. Bougas}
\affiliation{Department of Physics and LAMOR, Missouri University of Science and Technology, Rolla, Missouri 65409, USA}

\author{P.G. Kevrekidis}
\affiliation{Department of Mathematics and Statistics, University of Massachusetts Amherst, Amherst, Massachusetts 01003-4515, USA}

\author{C.-L. Hung}
\affiliation{Department of Physics and Astronomy, Purdue University, West Lafayette, Indiana 47907, USA}
\affiliation{Purdue Quantum Science and Engineering Institute, Purdue University, West Lafayette, Indiana 47907, USA}

\author{P. Schmelcher}
\affiliation{Center for Optical Quantum Technologies, Department of Physics, University of Hamburg,  Luruper Chaussee 149, D-22761 Hamburg, Germany}
\affiliation{The Hamburg Centre for Ultrafast Imaging, University of Hamburg, Luruper Chaussee 149, D-22761 Hamburg, Germany}

\author{S. I. Mistakidis}
\affiliation{Department of Physics and LAMOR, Missouri University of Science and Technology, Rolla, Missouri 65409, USA}

\date{\today}

\begin{abstract} 

We explore the correlated dynamics underlying the formation of the quantum Peregrine soliton, a prototypical rogue-wave excitation, utilizing interaction quenches from repulsive to attractive couplings in an ultracold bosonic gas confined in a one-dimensional box trap. 
The latter emulates the so-called semiclassical initial conditions
and the associated gradient catastrophe scenario facilitating the emergence of a high-density, doubly localized waveform. 
The ensuing multiorbital variant of the Peregrine soliton features notable deviations from its mean-field sibling, including a reduced peak amplitude, wider core,  absence of the side density dips, 
and earlier formation times. 
Moreover, Peregrine soliton generation yields coherence losses, while experiencing two-body bunching within each of its sides, which show anti-bunching between each other. 
Controllable seeding of the Peregrine soliton is also demonstrated by tuning the atom number or the box length, while reducing the  latter favors the generation of the time-periodic Kuznetsov-Ma breather. 
Our results highlight that correlations reshape the morphology of rogue waves in the genuinely quantum, nonintegrable realm, while setting the stage for the emergent field of quantum dispersive hydrodynamics. 

\end{abstract}

\maketitle
    
\section{Introduction}  \label{Sec:Introduction}

Rogue waves (RWs)--spatiotemporal highly localized waveforms 
that ``appear from nowhere and disappear without a trace''~\cite{akhmediev2009waves}, and are
characterized by amplitudes at least two times larger than their background--constitute an emerging class  of potentially disastrous, extreme nonlinear events~\cite{kharif2008rogue}. 
Their understanding is a topic of particular interest within the nonlinear wave community, drawing on observations from water tanks~\cite{Chabchoub_water,Chabchoub_water1}, plasmas~\cite{Bailung_plasmas}, nonlinear optics~\cite{kibler2010peregrine,solli2007optical},  oceanography~\cite{kharif2008rogue,perkins2006dashing}, and very recently ultracold quantum gases~\cite{Romero_experimental_2024}. 
Paradigmatic RW configurations bearing an analytical solution within the canonical integrable paradigm of the
nonlinear Schr{\"o}dinger (NLS) model with   attractive interactions are the following: (1) the celebrated Peregrine soliton (PS)~\cite{peregrine_water_1983}, which is a rational solution doubly localized in space and time, (2) the Akhmediev breather~\cite{akhmediev2009waves} being spatially periodic and
temporally localized, and (3) the Kuznetsov-Ma soliton~\cite{kuznetsov1977solitons,ma1979perturbed} featuring time periodicity along with space localization. There also exist extensions of these, such as (4) cnoidal-type structures~\cite{ChenPelinovsky2024,kuznetsov1974stability} corresponding to localized large amplitude humps mounted on top of periodic wave backgrounds.

A commonly expected to be necessary  (yet not clear whether it is sufficient) 
precursor~\cite{ruderman2010freak,baronio_vector_2014}  for RW formation is the modulational instability (MI) mechanism. This is  associated with the exponential growth of small perturbations on an attractively interacting unstable uniform background~\cite{zakharov_modulation_2009}.
Indeed, RW nucleation in NLS models featuring MI has been achieved by exploiting the gradient catastrophe mechanism developed by Bertola and Tovbis~\cite{bertola2013universality} for wave packets exhibiting
focusing behavior in the so-called semiclassical NLS regime.  Such
RWs have also emerged through utilizing interfering dam-break flows in the work of El, Khamis, and Tovbis~\cite{el2016dam}; see also Ref.~\cite{biondini2016universal}. 
The latter emanate from the well-established Riemann problem~\cite{el2016dam,gurevich1993modulational}, emulated by box (steplike) initial conditions and leading to the formation of counterpropagating dispersive shock wave (DSW) structures. Such DSW structures have been 
realized in various platforms~\cite{el2016dispersive}, including---but by no means limited to---nonlinear optics~\cite{wan2010diffraction} and ultracold atoms~\cite{Hoefer_DSW_observation,mossman2024nonlinear}.

Ultracold quantum gases, and, in particular Bose-Einstein condensates (BECs), have long been established, within the last 30 years, as
ideal systems to observe the above-discussed many-body (MB) nonlinear phenomena. 
This is due to their exquisite tunability in terms of the system parameters, such as the interparticle interactions via Feshbach resonances~\cite{chin2010feshbach}, and crafting arbitrarily shaped external potentials in different dimensions~\cite{bloch2008many,henderson2009experimental}. 
Notable examples, in the context of attractively interacting single-component BECs, include MI-assisted generation of bright solitary waves~\cite{strecker2002formation,nguyen2017formation,Robbins}, 
the study of their excitation spectrum and of higher-order variants
thereof ~\cite{Haller1},
the two-dimensional Townes soliton~\cite{Chen_observation_2020} and the necklace configurations thereof stemming from the destabilization of a vortex~\cite{banerjee2024collapse}, and the emulation of the dam-break flows yielding DSWs~\cite{tamura2025observation}. 
On the other hand, it can be shown that the description of a highly particle imbalanced two-component repulsively interacting BEC within the immiscible phase can be reduced to a single-component attractive Gross-Pitaevskii model for the minority component~\cite{Dutton_eff}.
Recently, this reduction scheme has been generalized to multicomponent settings~\cite{Bougas_vector_2025}.  
The above-mentioned effective  description has paved the way for the observation of the Townes soliton~\cite{Bakkali_realization_2021,bakkali_townes_2022}, the PS~\cite{Romero_experimental_2024}, and the nonlinear stage of MI~\cite{mossman2024nonlinear} in two-component immiscible BECs
(which can also lead to PS formation via the interference
mechanism of Ref.~\cite{el2016dam}). These multicomponent 
settings are amenable to the formation of vector PSs in 
two-~\cite{Baronio2012,baronio_vector_2014} and three-component settings~\cite{Bougas_vector_2025}.

The aforementioned phenomena have been studied within the mean-field (MF) approximation, while their MB  correlated character is, to the best of our knowledge, currently unexplored. 
It is indeed unclear whether and how the shape and properties of RW structures are affected when interparticle correlations are taken into account, especially as one departs from the integrable limit where 
the vast majority of studies have been limited, including in the ultracold atom realm; nevertheless, it is relevant to point out the presence of computationally identified extensions of the relevant waveforms, e.g., in Refs.~\cite{WardKevrekidisWhitaker2019,Charalampidis2018Rogue} and
more recently in the competing nonlinearity setting of Ref.~\cite{chandramouli2025rogue}.  
Importantly, an investigation of the beyond-MF features of
PSs is anticipated to shed light on the participating microscopic mechanisms of RWs through the building blocks of the MB  wave function and the ensuing correlation patterns.

Our study aims to provide an initial step toward the direction of quantum RWs by exploring the role of MB  effects in the experimentally relevant dynamical formation of RWs, utilizing the PS as a prototypical example. 
To nucleate the PS, we consider a one-dimensional (1D) bosonic gas confined in a box-potential with hard-wall boundary conditions. 
Starting from repulsive interactions where the bosons assemble in a nearly homogeneous density state, we follow a quench to attractive coupling strengths, a protocol that has been often leveraged in 
experimental practice from the early work of Ref.~\cite{Donley2001Bosenova}
and till the recent work of Ref.~\cite{tamura2025observation}. 
This process triggers the 
focusing of the original single-humped wave packet, partially also involving the emission of two counterpropagating density wave fronts from the box edges, which through their interference eventually give rise to  what we refer to  as the quantum variant of the PS.
It is also interesting to note that the considered parametric regime dictated by the atom number, box length, and postquench interactions is chosen such that MI of the background is suppressed in contrast to previous works~\cite{el2016dam,Adriazola_experimentally_2025,chandramouli2025rogue,tamura2025observation}.
To track the emergent nonequilibrium beyond-MF dynamics of the ultracold gas, we employ the multilayer multiconfiguration time-dependent Hartree method for atomic mixtures (ML-MCTDHX)~\cite{mlx_jcp_2013,cao2017unified,njp_mlx_2013}, and in particular its reduction to bosons, dubbed MCTDHB. 

We demonstrate that a high-density, doubly localized (in both space and time) RW structure representing the quantum variant of the PS is nucleated after the interference of the emitted density edges 
in the correlated dynamics of the considered non-integrable system
(which, however, bears an integrable MF limit variant). 
Importantly, our \textit{ab initio} MCTDHB computations reveal prominent correlation-induced deviations from the classical field prediction, ultimately modifying the PS morphology and dynamics.  
These alterations manifest as a reduced peak amplitude and wider core of the quantum PS, the absence of the characteristic side density dips
(a staple feature of the PS at the MF level) , and its earlier formation as compared to its MF  counterpart~\cite{Romero_experimental_2024}. 
These constitute the important correlation properties of the quantum PS, demonstrating its deviations from its known exact solution in the integrable limit.

These features are rooted in the significant occupation of higher-order natural orbitals in the MB wave function, and, in particular, in their spatially delocalized profile that is imprinted in the density of the PS. 
As a consequence, the bosonic gas features enhanced fragmentation and information entropy for larger postquench attractions. Turning to the correlation patterns of the quantum PS, we observe that it exhibits prominent coherence losses.
Additionally, its tails feature two-body bunching among themselves and anti-bunching between each other. 
We also explicate that the atom number and box length are suitable knobs for controllably engineering the PS generation, with smaller box lengths facilitating the creation of a time-periodic structure reminiscent of the Kuznetsov-Ma breather.

Our work unfolds as follows. In Sec.~\ref{Sec:Framework}, we introduce the MB Hamiltonian of the 1D box trapped Bose gas and specify experimentally relevant system parameters. 
Section~\ref{Sec:MB_approach} outlines the important aspects of the variational MB methodology, namely, the MCTDHB framework, used to capture the correlation properties of the PS. 
Section~\ref{Sec:Dynamics} elaborates on the dynamical generation of the quantum PS, analyzing its beyond-MF characteristics, ranging from the role of individual orbitals, the presence of fragmentation, and the impact of the system size, to the associated one- and two-body correlation patterns. 
In Sec.~\ref{Sec:Conclusions}, we summarize our findings and discuss future directions for exploring quantum  hydrodynamic phenomena in ultracold gases. 
Appendix~\ref{app:long_time} presents the long-time MB dynamics of the system where MI may be present and provides comparisons with the respective MF predictions. 
In Appendix~\ref{app:transition}, we delineate the transition from time-periodic breather solutions to the PS in the MF realm, while in Appendix~\ref{app:harmmonic_trap} we elucidate the impact of the harmonic trap in the emergent quench dynamics.
Appendix~\ref{app:converge} discusses the ingredients of the numerical simulations and examines their convergence.

\section{Bose Gas in a 1D box}\label{Sec:Framework}

We consider $N$ weakly interacting ultracold bosons of mass $m$, held by a 1D box potential of length $L$ along the $x$ direction. 
To achieve a 1D geometry, we further assume sufficiently strong  harmonic confinement along the perpendicular $y$, $z$ directions, e.g., characterized by $\sqrt{\omega_y \omega_z}=\omega_{\perp}$, where for simplicity $\omega_y=\omega_z \equiv \omega_{\perp}$. 
Accordingly, as long as $\hbar \omega_{\perp}$ is much  larger than all the other relevant energy scales, the atoms become kinematically constrained along the  elongated $x$ direction as is well known~\cite{pitaevskii2003bose} has also been demonstrated in recent 1D experiments for associated PS structures~\cite{Romero_experimental_2024,mossman2024nonlinear}.
Hence, a 1D description is adequate for addressing both the stationary and dynamical properties of the system. 
The MB Hamiltonian of this ultracold atom gas reads
\begin{equation}\label{hamiltonian_eq}
H = \sum_{i=1}^N \Bigg( - \frac{\hbar^2}{2m} \partial_{x_i} ^2 + V(x_i) \Bigg) + g \sum_{i<j}^N \delta(x_i-x_j). 
\end{equation}
Here, $x_i$ refers to the position of the $i$th particle, and $V(x_i)=0$ for $\abs{x_i} \leq L/2$, while $V(x_i)=\infty$ otherwise, represents the 1D box potential of length $L$. 
The potential facilitates the crafting of Riemann, namely, steplike initial conditions~\cite{el2016dam}, corresponding here to a density jump at the boundaries located at $x=\pm L/2$.  
Such initial conditions can nowadays be controllably engineered experimentally~\cite{mossman2024nonlinear,tamura2025observation,navon2021quantum,Tajik_designing_2019}. Upon quench to the attractive regime, 
they are expected to focus within the attractive medium, with the resulting interference leading to RW generation; see also Appendix~\ref{app:harmmonic_trap} for the effect of a harmonic trap on the dynamics. 
Here, the relatively weak repulsive interactions result in a large healing length for the considered box size, and therefore, the overall initial density is not completely uniform as it would be for larger repulsions and/or particle numbers. 
This prevents the full development of DSWs~\cite{el2016dispersive}, expected from a uniform density with a jump at the boundary that would emulate Riemann initial conditions.  
Yet, as we will argue below, our setting enables the unprecedented dynamical formation of a quantum PS structure.

At ultracold temperatures, $s$-wave scattering dominates and therefore the interparticle interactions between bosons are approximated in our 1D setting by a contact potential of  effective strength $g=2\hbar \omega_{\perp} a \left[ 1 - |\zeta(1/2)| \frac{a}{\alpha_{\perp}}  \right]^{-1}$~\cite{Olshanii_atomic_1998,Bergeman_atom_2003}.
Here, $a$ denotes the three-dimensional $s$-wave scattering length, $\zeta(\cdot)$ is the Riemann zeta function~\cite{Abramowitz_handbook_1948}, and $\alpha_{\perp}=\sqrt{\hbar /(m \omega_{\perp})}$ is the oscillator length in the perpendicular directions. 
The effective coupling strength can be arbitrarily tuned by means of Feshbach resonances~\cite{chin2010feshbach, feshbach2-RevModPhys.82.1225} or confinement-induced resonances~\cite{Olshanii_atomic_1998, confinement-resonance-2-PhysRevLett.97.193203}.
This tunability is crucial for the considered dynamical protocol, i.e., an interaction quench from initial repulsive couplings, $g_i$, to attractive interactions, $g_f$, aiming to spontaneously generate a quantum PS configuration.

Our setup can be experimentally realized, e.g., by employing a ${}^{133}$Cs BEC in a 1D box potential with $\omega_{\perp}= 2 \pi \times 500 ~ \rm{Hz}$, similarly to a recent experiment of Ref.~\cite{tamura2025observation}. In this case, the chemical potential, $\mu$, satisfies $\mu \ll \hbar \omega_{\perp}$. 
Throughout, we measure time, length, and interaction strengths in units of $\omega_{\perp}^{-1}$, $\alpha_{\perp}$, and $\sqrt{\hbar^3 \omega_{\perp}/m}$, respectively.
Moreover, we typically consider $N=20$ bosons trapped in a box of length $L=20$, while the prequench (postquench) interaction strength corresponds to $g_i=0.05$ ($g_f=-0.05$). Accordingly, in dimensional units $L \approx 8~\rm{\mu m}$ and the s-wave scattering length is quenched from $a\approx 180a_0$ to $-180a_0$ via a Feshbach resonance, where $a_0$ is the Bohr radius. The total propagation times on the order of $T=50$ correspond to $ \sim 16 ~ \rm{ms}$, well within reach in related recent experiments such as Ref.~\cite{tamura2025observation} (with the latter often extending to several seconds).

\section{Many-Body Approach and Wave function ansatz}\label{Sec:MB_approach}

To address the ground state of the MB system described by Eq.~\eqref{hamiltonian_eq} and subsequently monitor the quench-induced correlation dynamics, we employ the MCTDHB~\cite{mctdhb-alon-PhysRevA.77.033613,alon2007unified} variant of the generalized ML-MCTDHX~\cite{cao2017unified,mlx_jcp_2013,njp_mlx_2013} approach. 
This is a variational MB numerical method for the \textit{ab initio} solution of the time-dependent MB Schr{\"o}dinger equation. 
A central facet of this approach concerns the use of a variationally optimized time-dependent (comoving) basis. 
This time-dependent basis facilitates an efficient truncation
of the underlying Hilbert space as compared to MB methods using a time-independent basis, as it requires a relatively smaller basis size. 
Accordingly, this truncation scheme is tailored to capture the
entire system's correlations provided that numerical convergence is reached; see also Appendix~\ref{app:converge} for a more detailed discussion.
As such, it is possible to describe the role of correlations even for a mesoscopic number of interacting bosons, see, for instance, Refs.~\cite{Nguyen_parametric,katsimiga2017dark,mistakidis2018correlation}, while ensuring numerical convergence. 
Detailed discussions on the ingredients, applicability, and convergence of this approach for different ultracold atom settings ranging from single- to  multicomponent/spinor systems and cavities are provided in the recent reviews~\cite{Lode_review,mistakidis2023few} and references therein.

Specifically, the MB wave function is expanded in terms of time-dependent number states, $\ket{\vec{n} (t)}$, weighted by time-dependent expansion coefficients, $C_{\vec{n}}(t)$, as follows:
\begin{equation}\label{psi_ansatz_eq}
\Psi_{MB} = \sum_{\vec{n}|N} C_{\vec{n}}(t) |\vec{n} (t)\rangle.  
\end{equation}
In this expression, the summation runs over all ${N+M-1}\choose{M-1}$ possible number state configurations of $N$ bosons distributed in $M$ single-particle functions (SPFs), $\varphi_i(x,t)$. 
Additionally, the number state vector $\ket{\vec{n}(t)} = |n_1(t), \cdots, n_M(t) \rangle$ incorporates the occupation number $n_i(t)$ of the $i=1,2,\dots M$ SPF, 
subject to the total particle number constraint, i.e., $\sum_{i=1}^M n_i(t) =N$.
In the next step, each SPF is expanded into a time-independent discrete variable representation~\cite{light1985generalized} within a spatial grid characterized by  $\mathcal{M}$ points. 
For our simulations, we typically use $\mathcal{M}=250$ grid points in a spatial interval $[-10, 10]$, $N=20$ bosons, and $M=8$ SPFs ensuring numerical convergence; see also Appendix~\ref{app:converge}.

\begin{figure*}
\centering
\includegraphics[width=\linewidth]{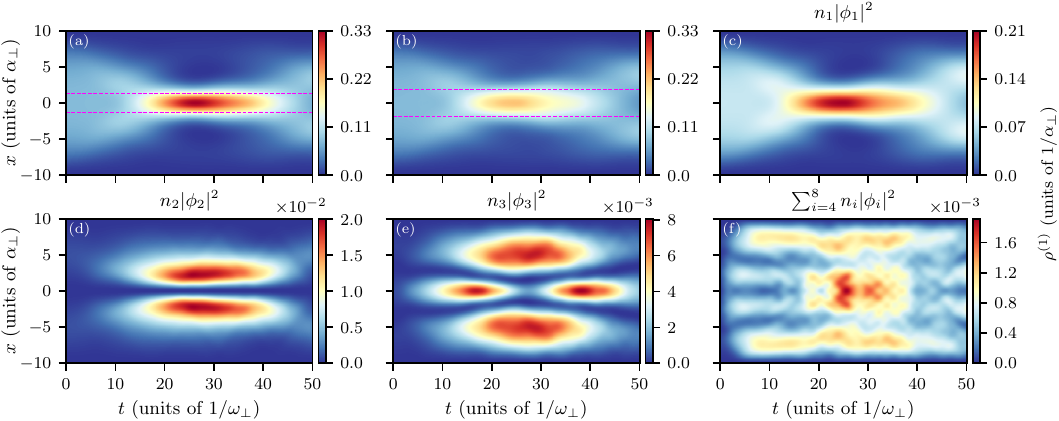}
\caption{Time evolution of the one-body density, $\rho^{(1)}(x,t)$, within (a) the  MF and (b) the MB approaches following an interaction quench from $g_i=0.05$ to $g_f=-0.05$. 
The quantum PS [panel (b)] forms at slightly earlier times and it is characterized by a relatively reduced (larger) 
peak amplitude (core) compared to its MF analog {as can also be seen in Figs.~\ref{fig3:density profile}(a) and ~\ref{fig3:density profile}(b)}. 
The magenta dashed lines in panels (a) and (b) provide a guide to the eye for the full width at half maximum of the PS core as predicted by the MF and the MB methods, respectively.
(c)-(f) Dynamics of the different orbital distributions, participating in the genuine MB evolution after the quench, weighted by their occupations, i.e., $n_i(t) \abs{\phi_i(x,t)}^2$ (see legends). 
The first orbital [panel (c)] resembles the MF configuration [panel (a)], while higher-order ones accommodate progressively more spatially delocalized structures. 
In all cases, the system consists of $N=20$ bosons trapped in a 1D box of length $L=20$, and it is prepared in its ground state with $g_i=0.05$.}
\label{fig1:1b-density}
\end{figure*}

The equations of motion for the coefficients, $C_{\vec{n}}(t)$, and the SPFs of the MCTDHB wave function ansatz are derived from a variational principle, e.g., the Dirac-Frenkel approach~\cite{frenkel1934wave}, determined by $\langle \delta \Psi |(i \hbar \partial_t - H) | \Psi \rangle = 0$. 
This process results in integro-differential equations for the expansion coefficients and SPFs~\cite{alon2007unified}. 
In order to compute the ground state of the repulsive gas, these equations are numerically solved using the standard  imaginary-time, $\tau = it$, propagation method.

In the limiting case of only one used SPF ($M=1$), the MCTDHB ansatz  reduces to the known MF product wave function, $\Psi_{MB}=\prod_{i=1}^{N} \Psi_{MF}(x_i,t)$, which ignores all interparticle correlations~\cite{pitaevskii2003bose}.   
Then, following a variational principle, we recover the standard single-component  Gross-Pitaevskii equation 
\begin{equation}
i\hbar \frac{\partial \Psi_{MF}}{\partial t}=\Big(-\frac{\hbar^2}{2m}\frac{\partial^2}{\partial x^2}  + V(x)+ g|\Psi_{MF} |^2 \Big) \Psi_{MF}.  
\end{equation}
In what follows, we systematically compare the formation of the quantum PS structure obtained within the MB approach with the predictions of the MF Gross-Pitaevskii approximation. 
This allows us to confirm the presence of the PS structure that has been discussed in the MF realm, but also explicate MB effects already evident at the density level, attesting unprecedented quantum PS signatures.

\section{Dynamical generation of the quantum PS}  \label{Sec:Dynamics}

As explained in the introduction, a promising dynamical protocol to seed RW nucleation is the quench from repulsive to attractive interactions, which we deploy herein. 
Within the MF approximation and for large atom numbers, Riemann-type initial conditions are expected to generate counterpropagating DSWs (known as dam-break flows~\cite{el2016dam}). These originate at the edges, travel toward the center, and interfere to produce RWs of various orders. 
This phenomenology has already been demonstrated in nonlinear optics~\cite{wan2010diffraction} and BEC~\cite{Adriazola_experimentally_2025,chandramouli2025rogue} setups.
An additional fundamental mechanism for the creation of PS
consists of the so-called Bertola-Tovbis scenario~\cite{bertola2013universality}, which starts
from a wide wave packet that focuses due to the attractive
nature of the effective nonlinear interaction (at the MF 
level of the Gross-Pitaevskii equation) and, through the
resulting gradient catastrophe at the focusing point, locally
forms PS structures. 
This is the predominant mechanism for PS generation in our setting,  largely suppressing the MI of the background, which would be otherwise prominent for larger atom numbers.

The key distinguishing feature of our work in comparison
to previous studies is that here we aim to investigate the currently elusive generation of the quantum variant of the PS and characterize its properties. 
To avoid complications stemming from the inherent MI of the attractive (postquench) background, the potential generation of higher-order 
RWs, and to ensure numerical convergence, we mainly study a low-atom-number setup with $N=20$.
For completeness, we remark that higher-order RWs were previously studied mathematically in Refs.~\cite{Bilman_2020,He_2013,bilman2} and were physically argued to be relevant to the MF BEC problem in Ref.~\cite{Adriazola_experimentally_2025}.
Notice, however, that due to $L=20$ the resulting density is experimentally detectable via \textit{in situ} measurements~\cite{frometa2025phase,wolswijk2025trapping,hung2015situ}. 
This choice naturally suppresses dam-break flow formation but allows the creation of the quantum PS due to the attractive (focusing) nature of the postquench setting. Indeed, as we will observe in what follows,
the relevant formation will bear features reminiscent of both the
Bertola-Tovbis gradient catastrophe scenario~\cite{bertola2013universality}
and ones of the El-Khamis-Tovbis DSW interference setting~\cite{el2016dam}
that have been previously presented in the nonlinear wave literature.
To assess the emergent quantum character of the PS, we rely on the MB quench dynamics of the bosonic system utilizing the MCTDHB method. 
As we argue below, this method gives access to the microscopic
character of the MB wave function but also the accompanied correlations determined through relevant entropies and coherence measures.

\subsection{Density evolution and interplay of orbitals}

\begin{figure}
\centering
\includegraphics[width=\linewidth]{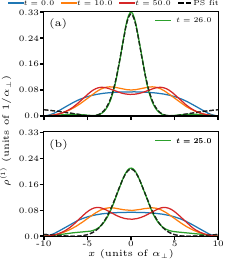}
\caption{Density snapshots (see legends) of the bosonic gas in the course of the interaction quench dynamics depicted in Fig.~\ref{fig1:1b-density}, as captured by  (a) the MF and (b) the MB methods. 
PS formation occurs at $t \sim 26$ ($t\sim 25$) in the MF (MB) evolution. 
The quantum PS edges are not fully dipped in contrast to the MF PS, and its peak amplitude (core) is reduced (increased) compared to its MF counterpart.   
Good agreement with the analytical PS solution at the integrable limit (see black dashed lines) takes place in the MF dynamics, while more prominent deviations are evident in the quantum variant. 
Other system parameters are the same as in Fig.~\ref{fig1:1b-density}.}
\label{fig3:density profile}
\end{figure}

The bosonic gas is prepared into its ground state with {weak} repulsive interactions of strength $g_i=0.05$. 
Hence, the underlying boson distribution corresponds to an almost homogeneous density profile with {relatively smooth} edges at the box boundaries [see Fig.~\ref{fig1:1b-density}(a) and ~\ref{fig3:density profile}(a)]. 
Following this preparation, we quench the interaction to relatively weak attractions, referring here to $g_f=-0.05$, and monitor the subsequent evolution of the system. 
{Such weak interaction strengths are chosen in order to achieve the fundamental PS generation, avoiding higher-order RW nucleation, as well as complex DSWs emanating from the edges preceding the PS formation. These are certainly intriguing phenomena that need to be explored in forthcoming studies.}

To visualize the dynamical response of the quenched bosonic gas, we first resort to the respective one-body reduced density matrix, whose spectral decomposition~\cite{Sakmann_RDM} reads
\begin{equation}\label{1-body-rdm-spectral}
\rho^{(1)}(x,x';t)=\sum_{i=1}^M n_i (t) \phi_i(x,t) \phi^*_i(x',t). 
\end{equation}
Here, $M=8$ denotes the number of considered SPFs in the MCTDHB expansion and $\phi_i(x,t)$ are the eigenfunctions of the one-body reduced density matrix being normalized to unity and called natural orbitals~\cite{Sakmann_RDM,roy2018phases}. 
The latter are related to the SPFs through a unitary transformation that diagonalizes $\rho^{(1)}(x,x';t)$ when expressed in the SPF basis; see also Refs.~\cite{mlx_jcp_2013,cao2017unified} for further details. 
The eigenvalues, $n_i(t)$, of $\rho^{(1)}(x,x';t)$ are dubbed natural populations taking values in the interval $n_i(t) \in [0, 1]$ and satisfying $\sum_{i=1}^{M} n_i(t) = 1$. 
They quantify the degree of the involved interparticle correlations by means that if $n_1(t)=1$ and $n_{i>1}(t)=0$, the system remains fully condensed, and the natural orbital $\phi_1(x,t)$ reduces to the MF wave function; see also Sec.~\ref{Sec:Frag}.

Specifically, we inspect the diagonal $\rho^{(1)}(x;t)=\rho^{(1)}(x,x'=x;t)$, which is the one-body density of the system and describes the spatial distribution of the atoms. 
This is readily tractable in experiments through an average over a sample of single-shot images~\cite{bloch2008many,wolswijk2025trapping}. 
After the quench to attractive interactions, here to $g_f = -0.05$, 
given the focusing nature of effective nonlinearity of the MF system, 
the atoms tend to contract toward the center ($x=0$) [see Figs.~\ref{fig1:1b-density}(a) and ~\ref{fig1:1b-density}(b)].  
This process is manifested through two counterpropagating density portions from the box edges  traveling to $x=0$ where they interfere. 
Recall that in the case of higher densities, i.e., larger atom number, the aforementioned moving density wave fronts form DSWs~\cite{hoefer2009dispersive,el2016dispersive}. 
The latter refer to highly spatially oscillatory waveforms with a monotonically varying wave envelope, which have been observed, for instance, in $^{87}$Rb experiments~\cite{Hoefer_DSW_observation,Chang_DSW}. 
These DSWs also need a wider spatial domain in order to fully form.
Moreover, the focusing phenomenology of Figs.~\ref{fig1:1b-density}(a) and ~\ref{fig1:1b-density}(b)
is also partially reminiscent of the gradient catastrophe scenario of Ref.~\cite{bertola2013universality}. In that light, we cannot fully 
attribute the PS-inducing interference to either the scenario of Ref.~\cite{el2016dam}
or that of~\cite{bertola2013universality}. 

Nevertheless, the interference of the above-discussed density fractions 
still produces a high-density structure around $x=0$, as identified within both the MF [Fig.~\ref{fig1:1b-density}(a)] and the correlated [Fig.~\ref{fig1:1b-density}(b)] approaches. 
Notably, the MF results predict a periodic recurrence of this localized structure (see also Appendix~\ref{app:long_time}) reminiscent of a time-periodic Kuznetsov-Ma breather in contrast to the correlated dynamics. It should be mentioned in this context that both the
scenario of Ref.~\cite{el2016dam} and that of Ref.~\cite{bertola2013universality} produce an array of local PS structures; however, in our narrow spatial domain,
this phenomenology cannot be fully attained (even at the MF level).
These local PS structures are only attained in the cases of larger box sizes and fixed $N$, or larger particle numbers and fixed $L$ (see also Appendix~\ref{app:transition}).
Moreover, soon after its formation the quantum PS  structure dissolves into a broad wave packet; see, e.g, Fig.~\ref{fig3:density profile}(b) at $t=50$ and Appendix~\ref{app:long_time} for the characteristics of the long-time evolution.

Before elaborating on the impact of correlations in the morphology of this spatially localized configuration, let us explain why it resembles a PS. 
To do so, we first focus on the MF predictions [Fig.~\ref{fig1:1b-density}(a)], which are easier to interpret in terms of classified RW analytical solutions (in the integrable limit) and in view of the arguably extensive previous RW studies in this context. 
As already mentioned, within our MF realm the resulting localized configuration appears to be (nearly) time periodic and hence it bears some similarity to the
well-known Kuznetsov-Ma breather~\cite{kuznetsov1977solitons,ma1979perturbed}. 
At the times where the latter reaches its maximum density, it is akin to a PS.
Importantly, at the time instant of its formation, this high amplitude waveform 
(of nearly three times the height of the original wave packet) 
is also characterized by side density nodes, see Fig.~\ref{fig3:density profile}(a) at $t=26$, that are further supportive of a PS configuration. 
To infer the presence of the latter, we perform a fit of our numerically obtained configuration to the analytical PS waveform~\cite{peregrine_water_1983} in the integrable limit,
\begin{equation}\label{peregrine_wf}
\Psi_P (x, t) = \sqrt{P_0} \Big[ 
1- \frac{4 \left(1+2i  \frac{t-t_0}{T_P}  \right) }{1+4 (\frac{x-x_0}{L_P})^2 + 4(\frac{t-t_0}{T_P})^2}
\Big] e^{i\frac{t-t_0}{T_P}}.
\end{equation}
Here, $L_P$ ($T_P$) represents the characteristic spatial (temporal) scale of the PS, $P_0$ designates the background density satisfying $T_P=L_P^2=1/ \sqrt{gP_0}$, while $t_0$ and $x_0$ refer to the time instant and position where the PS emerges. 
The latter are set to $x_0 = 0$ and $t_0=26$ according to the MF numerical results. 
A very good agreement can be observed to occur {\it locally} between the MF configuration and the density extracted from the analytical solution given by Eq.~(\ref{peregrine_wf}) [see the dashed line in Fig.~\ref{fig3:density profile}(a)]. 
Additionally, upon calculating the phase of the MF wave function it turns out that there is an almost $\pi$ phase jump (not shown) between the core and the side peaks of the observed structure further confirming the association with a PS. 

\begin{figure*}
\centering
\includegraphics[width=\linewidth]{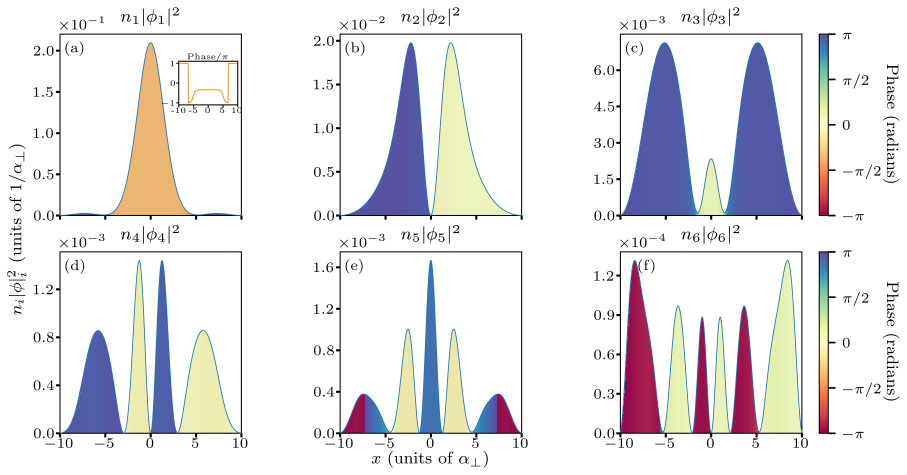}
\caption{(a)-(f) Density profiles of the various orbital distributions (see legend) at the time instant of the quantum PS formation, $t=25$. 
The first orbital is reminiscent of the MF density [see also Fig.~\ref{fig1:1b-density}(a)]. 
The inset in panel (a) shows the $\sim \pi$ phase jump between the core and the wings of the PS building atop the first orbital, which attests its nature.
The spatial delocalization of higher orbitals is responsible for the observed deviations from the MF PS predominantly enforcing finite tails, a wider core, and reduced peak amplitude of the quantum PS.  
The color shading in each orbital density configuration represents its phase. 
Almost $\sim \pi$ phase jumps occur between the density nodes appearing in the orbital densities.    
The remaining system parameters are the same as in Fig.~\ref{fig1:1b-density}.}
\label{fig6:orbitals_phase}
\end{figure*}

Turning to the MB predictions depicted in Figs.~\ref{fig1:1b-density}(b) and ~\ref{fig3:density profile}(b), we observe notable differences in the morphology of the generated spatiotemporal localized structure heralding the unprecedented---to our knowledge---genuinely quantum variant of the PS. Notice that 
we are not aware of any exact solution of the 
full MB correlated and nonintegrable system in 
closed (analytical) form. 
These correlation-induced deviations from the PS MF counterpart [Fig.~\ref{fig3:density profile}(a)] are manifested by a reduced peak amplitude, a broader PS core, and the absence of side density nodes [Fig.~\ref{fig3:density profile}(b)], with the quantum PS forming somewhat earlier. 
Namely, the high-density structure appears around $t \sim 25$ compared to its formation at $t \sim 26$ within the MF approach; see also the dynamics of the variance of the cloud [Eq.~(\ref{variance})] depicted in Fig.~\ref{fig5:varyN_L20_MB}, where the first minimum refers to the formation time of the quantum PS.

Using, for instance, a transverse confinement of $\omega_{\perp}=2 \pi \times 500~{\rm Hz}$ these times correspond to $7.96~{\rm ms}$, within the MB approach, and $8.28~{\rm ms}$, within the MF approach, respectively. Admittedly, this difference would be challenging to detect experimentally in sharp contrast to the width and peak amplitude of the quantum PS. 
Indeed, the width of the PS density distribution is $\sim 3.78$ ($\sim 2.65$) in dimensionless units within the MB (MF) method referring to $\sim 1.47~  {\rm \mu m}$ ($\sim 1.03 ~{\rm \mu m}$). 
The underlying width difference being on the order of $\sim 0.44~{\rm \mu m}$ can be experimentally resolved utilizing quantum gas microscopy fluorescence imaging~\cite{gemelke2009situ,gross2021quantum,Yao,Xiang,Jongh} or superresolution microscopy~\cite{McDonald,Subhankar_nanoscale_2019}. 
In a similar vein, the disagreement in the peak amplitudes of the PS density (normalized to unity) between the MF and the MB approaches is $\sim 0.12$, which refers to $\sim 0.29~{\rm \mu m^{-1}}$ and can also be experimentally assessed through the aforementioned quantum gas and superresolution  microscopy techniques.
These deviations are traced back to the presence of non-negligible two-body correlations within the MB case, as explained in  Sec.~\ref{Sec:Corr}. 
To emphasize the structural modifications of the quantum PS with respect to its MF counterpart, we carry out a fit of the analytical solution 
of Eq.~(\ref{peregrine_wf}) to the MB computed waveform at the time of  formation of the high-density structure; see the dashed line in Fig.~\ref{fig3:density profile}(b) at $t=25$. 
Although this fitting achieves good agreement with the quantum PS core, it fails to capture its tails due to the absence of the side density nodes when interparticle correlations are taken into account.

The observed alterations in the anatomy of the quantum PS, including the  absence of characteristic side density nodes that occur in the classical PS solution
(and the associated $\pi$ phase jumps), the wider core, and reduced amplitude [see  Fig.~\ref{fig3:density profile}(b) at $t=25$], are attributed to the superposition nature of the MB wave function. This is evident by the contribution and structure of the individual orbitals [see also Eq.~(\ref{1-body-rdm-spectral})]. 
In fact, as we will showcase later on, higher-lying orbitals and especially the second and third ones are significantly populated [see also Sec.~\ref{Sec:Frag} and Fig.~\ref{fig2:natural_populations}(a)]. 
To substantiate the impact of the different orbitals to the composition of the quantum PS, we subsequently analyze their spatial density structures ($|\phi_i (x,t)|^2$) during the dynamics. 
These are presented in Figs.~\ref{fig1:1b-density}(c)-\ref{fig1:1b-density}(f) where they are also scaled by the natural populations ($n_i$) such that their descending contribution for higher-order ones becomes clear from the colormaps.  
Starting with the first orbital, $|\phi_1 (x,t)|^2$, it is apparent that its overall evolution resembles the MF prediction [see Figs.~\ref{fig1:1b-density}(a) and ~\ref{fig1:1b-density}(c)]. 
Particularly, the profile of $|\phi_1 (x)|^2$ at $t=25$ along with its phase depicted in Fig.~\ref{fig6:orbitals_phase}(a) exemplifies that the peak amplitude is somewhat smaller and the core is shrunk compared to the quantum PS appearing in the single-particle density obtained within the MB method [Fig.~\ref{fig1:1b-density}(b)]. 
Moreover, there are side node density wings and an almost $\pi$ phase jump across the core and the side density peaks of the first orbital; see also the wave function phase of the first orbital in the inset of Fig.~\ref{fig6:orbitals_phase}(a). 

These similarities between the first orbital and the MF evolution suggest that the higher-order ones are responsible for the shape deformations of the quantum PS when contrasted to its MF counterpart. 
Similar structural deformations have been reported for the quantum analogs of solitary waves~\cite{Delande,katsimiga2017dark,Syrwid,Katsimiga_bent} and vortices~\cite{katsimiga2017many} where, for instance, their cores have been found to be wider compared to their MF siblings.  In those cases too, the higher orbitals
contributed to the loss of contrast of the MF-like behavior of the first orbital.
Indeed, it can be readily seen that the densities of the higher-lying orbitals, $|\phi_{i>1} (x,t)|^2$, are spatially extended [Figs.~\ref{fig1:1b-density}(d)-~\ref{fig1:1b-density}(f)] with the $i$th orbital exhibiting $i-1$ nodes accompanied by $\sim \pi$ phase jumps across them [Figs.~\ref{fig6:orbitals_phase}(b)-~\ref{fig6:orbitals_phase}(f)], a behavior that is reminiscent of higher-lying eigenstates. 
More concretely, the second [Fig.~\ref{fig1:1b-density}(d)] and third [Fig.~\ref{fig1:1b-density}(e)] orbitals (possessing the largest populations) show a notable delocalization tendency throughout the evolution. 
Here, $|\phi_{2} (x,t)|^2$ contributes significantly outside the PS core, while $|\phi_{3} (x,t)|^2$ exhibits a low-density core, surrounded by noticeable side density humps. 
These configurations are practically imprinted on the density of the gas and hence on the quantum PS, causing also a reduced peak amplitude and lifting the side density nodes of the first orbital.   
Similarly, even more delocalized distributions are building on top of the higher-orbital densities, as shown in Figs.~\ref{fig1:1b-density}(f) and ~\ref{fig6:orbitals_phase}(d)-(f), further contributing to the overall deviation of the quantum PS from its MF sibling.
Their effect becomes progressively weaker for higher orbital number, since their occupations decrease; see the following discussion and Fig.~\ref{fig2:natural_populations}(a).

\subsection{Dynamical fragmentation and entropies} \label{Sec:Frag}

\begin{figure}
\centering
\includegraphics[width=\linewidth]{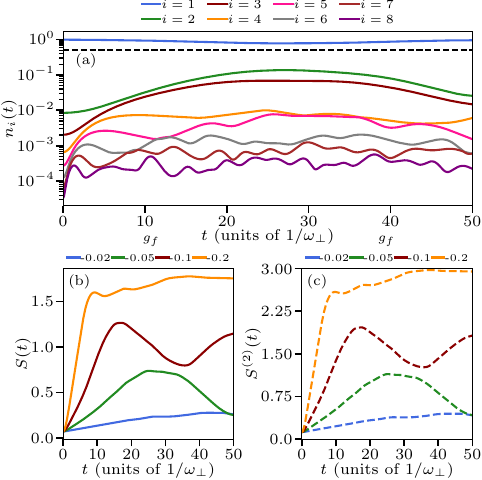}
\caption{(a) Time evolution of the natural populations of the eight orbitals participating in the MB wave function expansion. 
It is evident that the second and third orbitals are significantly occupied in the course of the evolution, while the remaining ones are mainly suppressed.  
The horizontal dashed line marks orbital population $n_{i}(t)=0.5$. 
Dynamics of the information entropy measures [Eq.~(\ref{Shannon})] are shown, namely, (b) $S(t)$ (solid lines) and (c) $S^{(2)}(t)$ (dashed lines), stemming from the one- and two-body reduced densities, respectively, for different postquench attractive interaction strengths $g_f$ (see legends).
As expected, quenches to stronger interactions yield a larger amount of fragmentation and hence degree of correlations. 
Other system parameters are the same as in Fig.~\ref{fig1:1b-density}. 
}
\label{fig2:natural_populations}
\end{figure}

The participation of multiple natural orbitals in the MB wave function [Eq.~(\ref{psi_ansatz_eq})]  alludes to the presence of interparticle correlations, yielding the phenomenon of fragmentation~\cite{Mueller_fragmentation}. 
This can be readily inferred by studying the involved natural populations, i.e., $n_i(t)$, being the eigenvalues of the one-body reduced density matrix [Eq.~\eqref{1-body-rdm-spectral}]. 
The time evolution of the individual orbitals used (here $M=8$) is illustrated in Fig.~\ref{fig2:natural_populations}(a) for the quenched gas of $N=20$ atoms from $g_i=0.05$ to $g_f=-0.05$.

Initially, $t=0$, at the ground state of the system with weak repulsive interactions there is negligible fragmentation on the order of $1-n_1(t) \leq 1.16\%$. 
This means that the prequench state is essentially condensed, namely, only minor deviations occur from an MF product state.  
This situation changes dramatically after the quench to attractive couplings. 
Indeed, the second, $n_2(t)$, and third, $n_3(t)$, natural populations are systematically increasing in the course of the evolution, featuring a maximum value [accompanied by the largest reduction of $n_1(t)$] within the interval of PS formation. 
The {horizontal} dashed line 
at $0.5$ in Fig.~\ref{fig2:natural_populations}(a) allows one to clearly visualize the substantial increase of $n_2(t)$ and  $n_3(t)$. 
Afterward, $n_2(t), n_3(t)$ show a descending tendency while retaining their significant contribution in the course of the MB dynamics. 
The higher-lying orbitals exhibit a similar to the above-described trend but they have substantially smaller populations, implying a diminishing contribution to the MB state. 
As a result, a high degree of fragmentation takes place becoming maximum close to the emergence of quantum PS implying its inherent multiorbital character (and thus 
departure from the well-established MF limit). 
This increase of correlations at the PS formation is attributed to the arguably  prominent focusing tendency of the atoms when they assemble in this spatiotemporally localized structure.  

Next, we quantify the overall degree of Hilbert space fragmentation and correlation spreading during the quantum PS generation. 
For this purpose, we first compute the natural population Shannon information entropy measure~\cite{roy2018phases}, which is capable of quantifying the multi-orbital character of the quantum PS, 
\begin{equation}
S(t)=-\sum_{i=1}^M n_i(t) \ln[n_i(t)].\label{Shannon}  
\end{equation}
It is apparent that for a MF product state $S(t)=0$, while for actual multiorbital configurations (such as the quantum PS) it holds that $S(t) > 0$ since $n_{i>1}(t) \neq 0$. 
The time evolution of $S(t)$ is depicted in Fig.~\ref{fig2:natural_populations}(b) for several post quench interaction strengths ($g_f$) in the system with $N=20$ and $L=20$. 
We observe that in all cases, $S(t)$ gradually increases as the two density fractions, emanating from the box edges travel toward $x=0$. 
The relevant quantity features a maximum (e.g., {$S(t\approx 17) \simeq 1.26$ for $g_f=-0.1$}) during their interference, where the quantum PS forms, and then slightly reduces as the gas distribution delocalizes [Fig.~\ref{fig1:1b-density}(b)].  
This behavior implies that the quantum PS is an almost  maximally fragmented state. 

Notice here that a maximally (or $M$-fold) fragmented MB state refers to the situation where the information entropy becomes maximal, i.e., $S_{{\rm max}}(t)=\ln(M)$ and $n_i(t)=1/M$, with $M$ denoting the used number of SPFs dictating the truncation of the MB wave function [see also Eqs.~(\ref{psi_ansatz_eq}) and ~(\ref{1-body-rdm-spectral})]. 
In our case, $S_{{\rm max}}(t) \approx 2.08$. 
Finally, it can be seen that $S(t)$ is larger for increasing magnitudes of the postquench attraction, which means that the role of correlations becomes, as expected, more pronounced.

As a complementary investigation on the multiorbital character of the quantum PS and its accompanied correlation spreading, we study the two-body information entropy, $S^{(2)}(t)=-\sum_{i}^{M(M+1)/2} n^{(2)}_i(t) \ln[n^{(2)}_i(t)]$, constructed from the eigenvalues of the two-body reduced density matrix. 
The spectral decomposition of the latter reads 
\begin{equation}
\begin{split}
\rho^{(2)}(x_1, x_2 | x_1', x_2';t)
 &= \sum_{i=1}^{M(M+1)/2} n^{(2)}_i(t) \\
 &\quad \times \alpha^{(2)}_i(x_1,x_2;t)\,
          \alpha^{*(2)}_i(x_1',x_2';t).
\end{split}
\label{eq:2b_spectral}
\end{equation}
where $n^{(2)}_i(t)$ and $\alpha^{(2)}_i (x_1,x_2;t)$ are the eigenvalues and eigenfunctions (also known as natural geminals) of $\rho^{(2)}(x_1, x_2 | x_1', x_2';t)$, respectively~\cite{Sakmann_RDM}. 
The diagonal ($x_1=x'_1 \equiv x$ and $x_2=x'_2\equiv x'$) of the two-body reduced density matrix, $\rho^{(2)}(x, x';t)$, provides the probability distribution of measuring simultaneously two particles located at positions $x$, and $x'$, respectively.  
The case of $S^{(2)}(t)=0$ is associated with a pure two-body density matrix, which in turn implies absence of two-body correlations. 

The time evolution of the two-body information entropy after the interaction quench is presented in Fig.~\ref{fig2:natural_populations}(c), displaying a similar trend
with the respective $S(t)$ [Eq.~\eqref{Shannon}].
It turns out that irrespective of $g_f$, the two-body entropy is non zero and reaches its maximum near the PS formation, e.g., $S^{(2)}_{\rm{max}} (t\approx 17.4) \simeq 1.96$ for $g_f=-0.1$, where two-body correlations become significant. 
Importantly, the two-body information entropy remains below the upper bound set by our Hilbert space truncation, i.e. $S^{(2)}_{\rm{max}}(t)=\ln (M(M+1)/2) \approx 3.58$. 
This further hints that the truncation in the MB ansatz is sufficient to describe the quantum PS, and that the two-body reduced density matrix is not maximally mixed. 
We remark that the evaluation of entropy measures during RW formation is certainly an intriguing prospect for further future investigations, not only in single-component gases but also in multicomponent ones, in order to infer state preparation and entanglement propagation. It is worthwhile to mention here that while multicomponent
PS analogs have been devised at the MF level in the work
of Ref.~\cite{Baronio2012} for attractive and Ref.~\cite{baronio_vector_2014} for repulsive
multicomponent gases, we are not aware of relevant considerations at
the MB level.

\begin{figure*}
\centering
\includegraphics[width=\linewidth]{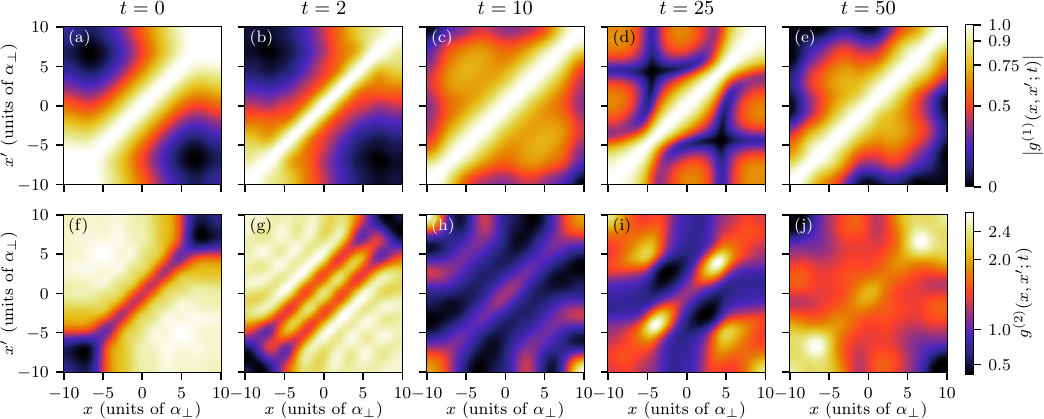}
\caption{Profiles of (a)-(e) the one-body and (f)-(j) the two-body coherence functions at different time instants (see legends) of the MB evolution obtained with MCTDHB. 
The quantum PS structure is characterized by loss of one-body coherence between its edges, see off-diagonals in panel (d), while it features a strong bunching effect within its left and right parts as well as anti-bunching between them [panel (i)]. 
System parameters are the ones used in Fig.~\ref{fig1:1b-density}. 
}
\label{fig4:coherence functions}
\end{figure*}

\subsection{Correlation patterns of the Peregrine}  \label{Sec:Corr}

As already discussed in Sec.~\ref{Sec:Frag}, the non-negligible contribution of the natural orbitals quantified through $n_{i>1}(t) \neq 0$ signifies the involvement of interparticle correlations during the RW formation. 
To identify the degree of these correlations arising in the quench-induced dynamics at the one-body level, we employ the normalized spatial first-order coherence function~\cite{Naraschewski}
\begin{align}\label{coh1}
g^{(1)} (x, x';t) = \frac{\rho^{(1)} (x, x';t)}
{\sqrt{\rho^{(1)}(x;t) \rho^{(1)}(x';t)}}.
\end{align}
This diagnostic essentially measures the deviation of the MB state from a MF product state for a fixed set of coordinates $x$, $x'$. 
Specifically, $\abs{g^{(1)} (x, x';t)}$ takes values within the interval $[0, 1]$. 
In the above expression, $\rho^{(1)} (x, x';t)$ is the one-body reduced density matrix of the system [see Eq.~(\ref{1-body-rdm-spectral})], whose diagonal is the single-particle density, namely, $\rho^{(1)} (x;t)=\rho^{(1)} (x, x'=x;t)$. 
Two different non-overlapping spatial regions $D_1$, $D_2$, where $x \in D_1$ and $x' \in D_2$, are characterized as fully coherent (incoherent) when $g^{(1)} (x, x';t)=1$ ($g^{(1)} (x, x';t)=0$).    
Accordingly, full coherence means that the MF state ignoring all correlations is sufficient to describe the system, while $\abs{g^{(1)} (x, x';t)}<1$ implies that two different spatial regions are partially incoherent and hence interparticle correlations play a nontrivial role.

At $t=0$, where the bosonic gas is in its ground state with relatively  weak repulsive interactions ($g_i=0.05$), coherence losses are observed across the off-diagonal of $\abs{g^{(1)} (x, x \neq x';t=0)}<1$ [Fig.~\ref{fig4:coherence functions}(a)].  This means that quasi-long-range order is progressively destroyed at longer distances from $x=0$ with the regions close to the box edges, i.e., $x=-x' \in [\pm 10,\pm 5]$, being nearly fully incoherent with one another. 
After the quench to the attractive regime, at early evolution times (e.g., $t=2$), where slight deformations occur at the edges of the original homogeneous state [Fig.~\ref{fig1:1b-density}(b)], the aforementioned coherence pattern is maintained [Fig.~\ref{fig4:coherence functions}(b)], with off-diagonal incoherent regions expanding to the center and shrinking the fully coherent diagonal region. 
Subsequently, e.g., around $t=10$, coherence exhibits a partial revival, especially along the off-diagonal of $\abs{g^{(1)} (x, x \neq x';t=10)}$ [Fig.~\ref{fig4:coherence functions}(c)], corresponding to the time interval where the emitted density wave fronts approach each other [Figs.~\ref{fig1:1b-density}(b) and ~\ref{fig3:density profile}(b)]. Turning to the time instance of the Peregrine formation, i.e., at $t=25$ presented in Fig.~\ref{fig4:coherence functions}(d), we observe that while the PS core is fully coherent, the two spatial regions in its vicinity (meaning the PS edges) feature substantial incoherence between each other, e.g., $\abs{g^{(1)} (x=5,x'=-5;t=25)} \ll 1$. 
This unprecedented correlation pattern of the quantum PS testifies the correlated  character of the emergent high-density structure depicted in Fig.~\ref{fig3:density profile}(b) at $t=25$. 
At longer evolution times, coherence may again be partially restored, as evidenced in Fig.~\ref{fig4:coherence functions}(e).  
We attribute this partial coherence recurrence during the evolution to the {weakly interacting nature of the} finite-size system, facilitating the continuous interference of the bosonic gas (see also Appendix~\ref{app:long_time}). 
Accordingly, the atoms reassemble close to their initial configuration [e.g., Fig.~\ref{fig3:density profile}(b) at $t=50$], leading to coherence patterns similar to the original one.

The correlated nature of the quantum PS structure is next explored at the two-body level, by estimating the degree of two-body correlations through the two-body coherence function~\cite{Sakmann_RDM,mistakidis2018correlation}, defined as
\begin{equation}\label{coh2}
g^{(2)} (x, x';t) = \frac{\rho^{(2)}(x, x';t)}{\rho^{(1)} (x;t) \rho^{(1)}(x';t)}, 
\end{equation}
with $\rho^{(2)}(x, x';t)$ representing the diagonal of the two-body reduced density matrix [Eq.~(\ref{eq:2b_spectral})].   
A fully condensed MB state is associated with $g^{(2)} (x, x';t)=1$, and implies that the two bosons at $x$ and $x'$ are two-body uncorrelated. 
In contrast, if $g^{(2)} (x, x';t)>1$ [$g^{(2)} (x, x';t)<1$], the MB state is termed two-body correlated (anti-correlated) describing the bunching (anti-bunching) tendency of two bosons residing in the respective spatial regions~\cite{mistakidis2023few}. 
The second-order coherence function is experimentally tractable using \textit{in situ} density fluctuation measurements~\cite{hodgman2011direct,tavares2017matter,Nguyen_parametric}. 

\begin{figure*}
\centering
\includegraphics[width=\linewidth]{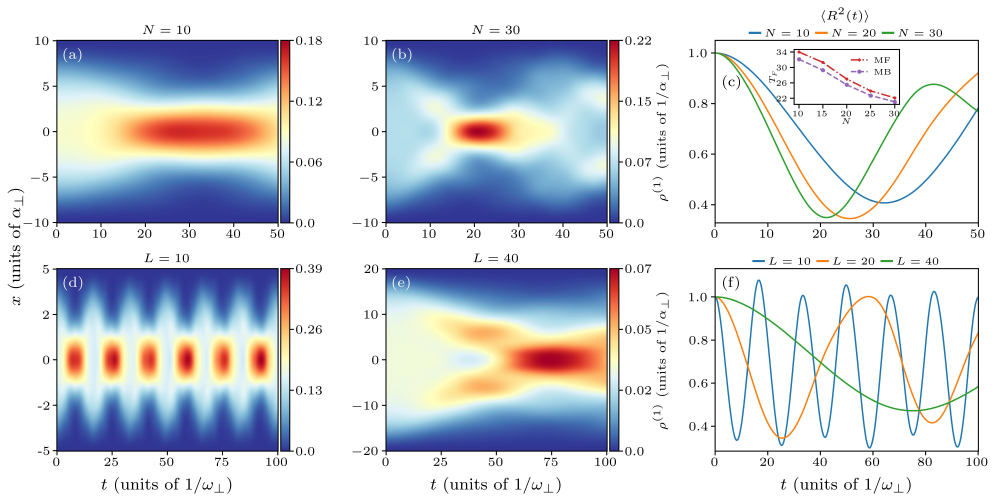}
\caption{Spatiotemporal dynamics of the one-body density, $\rho^{(1)}(x,t)$, within the MB approach [panels (a) and (b)] for different number of bosons $N$ (see legends) and fixed box length $L=20$ as well as [panels (d) and (e)] for various $L$ (see legends) and $N=20$. 
Time evolution of the rescaled position variance, $\braket{R^2(t)}=\braket{X^2(t)}/\braket{X^2(0)}$, for (c) varying $N$ and $L=20$ and (f) distinct $L$ keeping $N=20$ (see legends). 
The first minimum of the variance quantifies the emergence of the quantum PS, which manifests faster for either larger $N$ and fixed $L$ or smaller $L$ and constant $N$. 
In the latter case, a transition from a quantum PS to a quantum Kuznetsov-Ma breather occurs.
Inset in panel (c) presents the PS formation time, $T_F$, with respect to varying atom number in both the MF and MB methods. 
In all cases, the bosonic gas is initiated into its ground state with $g_i=0.05$ and a quench to $g_f=-0.05$ is considered.
}
\label{fig5:varyN_L20_MB}
\end{figure*}

The ground state of the weakly repulsive bosonic gas is characterized by an overall bunching behavior; see both the diagonal and off-diagonal elements of $g^{(2)} (x, x';t=0)$ in Fig.~\ref{fig4:coherence functions}(f). 
Moreover, the degree of bunching is enhanced for two bosons residing on opposite sides of the box potential, a behavior that is attributed to the repulsive interactions. 
Near the box edges, where the density is substantially depleted due to the boundaries, an anti-bunching tendency takes place; see, e.g., $g^{(2)} (x=9, x'=9;t=0) \to 0.5$. 
After the quench to attractive interactions, the two density fractions from the edges travel toward $x=0$ [Fig.~\ref{fig1:1b-density}(b)], causing a progressive modification of the above-discussed two-body correlation pattern [Figs.~\ref{fig4:coherence functions}(g) and ~\ref{fig4:coherence functions}(h)]. 
Specifically, at short times (e.g., $t=2$) the degree of bunching increases, as indicated by $g^{(2)} (x, x';t)$, deviating further from unity [Fig.~\ref{fig4:coherence functions}(g)], a trend that is traced back to the attractive postquench  interactions. 
However, as the counterpropagating density humps tend closer to $x=0$ around $t=10$ [Fig.~\ref{fig3:density profile}(b)], two-body bunching within each wave front reduces as can be deduced from the diagonal of $g^{(2)} (x, x'=x;t=10)$ [Fig.~\ref{fig4:coherence functions}(h)]. 
However, anti-bunching builds up between the edges of the same wave front or different ones.

At the quantum PS formation, a strong two-body correlated behavior occurs across the PS  core and within each of its sides; see the diagonal of $g^{(2)} (x,x;t=25)$ with $ \abs{x} \leq 4.5$ in Fig.~\ref{fig4:coherence functions}(i). 
This strong bunching effect stems from the contracting tendency of the gas forming a spatially localized density structure due to the underlying attractive interactions. This effect is also responsible for the earlier formation of the PS in the MB case, as compared to the MF scenario.  
Moreover, two particles positioned at opposite sides of the quantum PS feature an anti-correlated behavior, as can be deduced from the off-diagonal elements (e.g., at $x=-x' \sim 2.5$) near the PS core. Interestingly, bunching takes place among the spatial regions of the PS tails (e.g., $x=5$, $x'=-4.5$), 
having suppressed density.  
After the PS dissolves giving its place to a spatially delocalized distribution, an overall two-body bunching is reestablished with the two density humps moving outward to the box edges [Fig.~\ref{fig3:density profile}(b)], displaying a strongly correlated behavior [Fig.~\ref{fig4:coherence functions}(j)].

\subsection{Controllable seed of the quantum PS and transition to the Kuznetsov-Ma}\label{sec:KZ_toPS_trans}

Having analyzed the correlation properties of the quantum PS after an interaction quench to attractive interactions, we next elaborate on experimentally accessible knobs for its realization. 
In particular, we focus on quenches from $g_i=0.05$ to $g_f = -0.05$, and consider variations of the atom number, $N$ (box length, $L$), while keeping constant the box length $L$ (atom number $N$). 
Recall here that the box characteristics, including its length, can be experimentally adjusted using digital micromirror devices~\cite{navon2021quantum,gauthier2016direct}, while atomic samples with different atom number can also be  realized~\cite{wenz2013few,Lester}.

The time evolution of the gas one-body density, $\rho^{(1)}(x,t)$, after the  interaction quench for distinct atom numbers $N=10,30$ and fixed $L=20$  is illustrated in Figs.~\ref{fig5:varyN_L20_MB}(a) and ~\ref{fig5:varyN_L20_MB}(b); see also Appendix~\ref{app:transition} for systems with larger $N$. 
As $N$ increases (implying that the effective nonlinearity of the system
increases), the emission of counterpropagating density wave fronts from the box edges accelerates, leading to faster interference and the earlier formation of the quantum PS. 
Recall that larger $N$ implies smaller healing length of the initial density, and hence larger kinetic energy of the density wave fronts emanating from the edges, subsequently leading to reduced timescales of the PS formation.
We note that a similar acceleration takes place within the MF approach but with larger (smaller) Peregrine amplitudes (widths), presence of side nodes that are absent in the MB case, and slightly slower PS formation times compared to the correlated scenario; see also the inset of Fig.~\ref{fig5:varyN_L20_MB}(c).  
Additionally, preamble signatures of MI, associated with the
formation of localized density waveforms, following the quantum PS generation are evident for $N=30$ at $t>40$ where two major density humps appear. 
The long-time evolution of our {main} setup is discussed in more detail in Appendix~\ref{app:long_time}, where for this $g_f$ recurrence phenomena
occur for longer times, while for more negative $g_f$, persistent spatial localization at the level of the single-particle
density arises.

A suitable observable for capturing the PS formation time and the overall collective dynamics of the quenched bosonic cloud is the position variance, experimentally tractable via \textit{in situ} absorption imaging~\cite{Ronzheimer_exp,hung2015situ}, 
\begin{equation}
\braket{X^2(t)}=\braket{\Psi_{MB}(t)|x^2|\Psi_{MB}(t)}.\label{variance} 
\end{equation}
It can be employed in our setup to identify the onset of PS formation, since its first minimum corresponds to the largest cloud contraction, practically coinciding
with the PS creation. 
Moreover, it can also be used as a probe for the presence of beyond-MF correlations during the dynamics (see also Appendix~\ref{app:long_time}). 
Without loss of generality, we monitor the rescaled (to the original time) 
position variance, i.e., $\braket{R^2(t)}=\braket{X^2(t)}/\braket{X^2(0)}$, which is illustrated in Fig.~\ref{fig5:varyN_L20_MB}(c). 
We deduce that the PS nucleation as identified by the first minimum of $\braket{R^2(t)}$ is faster for larger $N$ while $L$ is held constant, as well as that the PS amplitude is reduced for smaller $N$. 
The time of the PS development, $T_F$, in both the MB and the MF frameworks is indeed in general smaller for increasing $N$ as shown in the inset of Fig.~\ref{fig5:varyN_L20_MB}(c). 
An essentially monotonic decrease of $T_F$ for increasing $N$ is observed, with $T_F$ being somewhat larger within the MF. 
The development of two-body correlations leads to a consistently smaller PS formation time within the MB approach (in comparison to the MF case). Nevertheless, the MF scenario helps to intuitively explain the smaller PS formation time for larger $N$,
given the associated stronger effective nonlinear interaction.

Next, we examine the impact of the box length $L$, while keeping the particle number fixed. 
In practice, a larger $L$ implies reduction of the initial density, together with its spatial width. 
It is not possible to understand the resulting dynamics by solely relying on the $N$ variation argument discussed above. Instead, the dynamics motivates analogies with earlier studies on how the width of Gaussian initial conditions influences the spontaneous nucleation of different RW structures~\cite{Charalampidis2018Rogue}, based on the semiclassical limit of the NLS model. 
Strikingly, it turns out that for decreasing $L$, and thus reducing the  width of the nearly homogeneous initial state, a time-periodic spatially localized configuration emerges in the course of the evolution of almost constant maximum amplitude; see also the underlying $\braket{R^2(t)}$ in Fig.~\ref{fig5:varyN_L20_MB}(f).  
This hints at the creation of a structure more reminiscent of 
time periodicity of the Kuznetsov-Ma breather~\cite{kuznetsov1977solitons,karjanto2021peregrine}, rather than a quantum PS. 
As argued in the work of Ref.~\cite{Charalampidis2018Rogue}, upon suitable decrease
of the width, the outcome of the relevant MF computations was more breather-like
(and eventually solitonic) in nature, being reminiscent of the classic 
Satsuma-Yajima work on higher-order MF solitonic structures~\cite{SatsumaYajima1974}. 
Indeed, the corresponding structure can be equally generated within the MF method (not shown for brevity), and it is characterized by a fixed maximum amplitude and a
nearly periodic behavior in time; see also Appendix~\ref{app:transition} for the MF transition region across the $N$-$L$ plane between the periodic breather and the PS. 
We remark that the respective formation times and peak amplitude of the 
resulting breathing waveform are both slightly larger within the MF. 
Moreover, another intriguing observation stemming from the computation of the natural orbitals is that the degree of correlations or fragmentation identified through either $1-n_1(t)$ or the information entropy, $S(t)$, is smaller for this structure
in comparison to the PS.

On the other hand, larger box lengths such as $L=20,40$ result in the dynamical creation of the quantum PS as shown in the relevant density dynamics of  Fig.~\ref{fig5:varyN_L20_MB}(e) and the position variance dynamics of Fig.~\ref{fig5:varyN_L20_MB}(f). 
A larger box length, e.g., $L=40$, yields slower formation times than smaller boxes, since the emitted density fractions from the box edges need to travel a larger distance
(for the same value of $N$). 
This phenomenology has been confirmed for even larger box sizes, e.g., $L=100, 200$ (not shown for brevity) within the MF approximation. The latter settings lead to the
more pronounced development of MI
features, as well as the emergence of recurrence 
phenomena at considerably longer timescales, as
imposed by the respective domain sizes.
Also, smaller peak amplitudes occur  due to the decreased initial atom density. 
Summarizing, we have demonstrated a transition from the quantum PS to a breathing waveform for decreasing box length and fixed atom number, but also exemplified that both $N$ and $L$ can be utilized to control the time of formation and peak amplitude of RW structures.

\section{Conclusions and Perspectives}\label{Sec:Conclusions}

We have investigated the unprecedented dynamical generation of the 
quantum Peregrine soliton in a correlated 1D Bose gas trapped in a box potential. This potential is designed to engineer 
initial conditions that via focusing (in the attractive regime) and interference
may produce a quantum analog of the PS pattern. 
To achieve a complete quantum description of the ensuing focusing evolution, we rely on the \textit{ab initio} MCTDHB method, which allows to capture the interparticle correlations of the system and hence operate beyond the MF approximation. 

Starting from repulsive interactions, where the Bose gas possesses a nearly uniform,
chiefly condensed distribution, we devise quenches to attractive interactions. Such a protocol favors the emergence of two counterpropagating density wave fronts from the edges moving toward the center where they collide and interfere. 
As a result, a transient doubly localized in space and time 
structure that we refer to as the quantum variant of the PS emerges. 
To testify the PS nature of the nucleated configuration and uncover its quantum features, we also simulate the corresponding MF dynamics, revealing the formation of a high-amplitude time-periodic configuration partially reminiscent of the Kuznetsov-Ma breather. 
These localized MF  solutions are compared to the analytically available in the integrable case PS waveform, unveiling reasonable agreement. 
Moreover, a phase jump is observed between the core and the side peaks of the PS
in the MF limit, a feature that is smeared out by higher-order orbitals
in the quantum case.

Accordingly, we explicate how the morphology of the quantum variant of the PS excitation deviates markedly from its MF counterpart known to occur within the respective focusing NLS. 
The quantum PS forms at relatively earlier times, possesses a significantly reduced peak amplitude and wider core, and lacks the characteristic side density dips accompanying its MF sibling. 
The aforementioned deviations arise from the dynamical build up of interparticle correlations, identified by the prominent occupations of multiple natural orbitals. 
The latter exhibit (progressively more so for higher orbitals) delocalized density distributions, resulting in the reshaping of the MF PS whose characteristics are mainly captured by the first orbital. 
This established multiorbital character of the PS yields pronounced fragmentation, especially close to the time interval of its formation. 
Inspecting the associated correlation functions, we reveal intriguing patterns. 
Namely, coherence is lost at the edges of the PS density peak, while two-body bunching occurs inside the PS core and prominent anti-bunching takes place between its symmetric spatial regions.

Finally, we examine the impact of the atom number and the box size on the PS nucleation in order to regulate its manifestation through experimentally accessible parameters. 
We have quantified the intuitive expectation that increasing the number of bosons (box length) with fixed box length (atom number) accelerates (slows down) the appearance of the correlated peak structure. 
Importantly, decreasing the box length with constant particle number favors the nucleation of a time-periodic structure, partially 
reminiscent of the Kuznetsov-Ma breather (and partially also of the
fundamental Satsuma-Yajima realization of higher order solitons) within the MB approach.

Our results set the stage for an emergent, yet so far largely unexplored, research field of {\it quantum dispersive hydrodynamics}, motivating a plethora of highly compelling future studies aiming to probe the interplay between integrability, dimensionality, and MB physics. 
A natural next step is to study the quench dynamics to stronger attractive interactions to generate complex RW lattice structures, such as the Christmas tree known to occur within MF  theory~\cite{Charalampidis2018Rogue}. 
Another important open question already at the MF  level is to explore the persistence of the PS, and in general, of RWs in the dimensional crossover, appreciating the role of transverse excitations. 
Moreover, it is interesting to consider an initially harmonically confined gas and follow a double quench protocol releasing the trap and simultaneously changing the interactions from repulsive to attractive for nucleating higher-order RWs (HORWs) as in the recent work of Ref.~\cite{Adriazola_experimentally_2025}. It is an open question whether MB 
analogs of such---more delicate---HORWs may exist, all the way to the
so-called infinite-order RW of Ref.~\cite{Suleimanov2017}.
Certainly, the study of other types of hydrodynamic excitations in the quantum realm such as the DSWs~\cite{el2016dispersive} is desirable, stemming from
a MB variant of the celebrated Riemann problem.
A first step in this direction can be achieved by engineering suitable boxlike initial densities and analyzing their evolution when embedded to the quantum MB environment. 
Finally, as was indicated explicitly earlier in this work, generalizations to two- (and more generally larger) component systems for both RWs and shock waves offer highly intriguing directions toward quantum MB generalizations.

\begin{acknowledgements}

The authors are grateful to P. Engels and S. Mossman for stimulating discussions on the topic of rogue waves. 
S.I.M. acknowledges support by the Army Research Office under Award No. W911NF-26-1-A043.
D.D. and P.S. acknowledge funding by the Cluster of Excellence “CUI: Advanced Imaging of Matter” of the Deutsche Forschungsgemeinschaft (DFG)---EXC 2056---Project ID 390715994.
P.G.K. was supported by the U.S. National Science Foundation under the Awards No. PHY-2110030, No. DMS-2204702, and No. PHY-2408988 and  
a grant from the Simons Foundation SFI-MPS-SFM-
00011048. C.-L.H. acknowledges support from the NSF (PHY-2409591) and the AFOSR (FA9550-22-1-0327).
The numerical computations were performed using the PHYSnet computational cluster at the University of Hamburg. D.D. gratefully acknowledges the technical support of Martin Stieben and Dr.~Tomohiro Hashizume.

\end{acknowledgements}

\appendix

\section{MI signatures in the long-time correlated  dynamics}\label{app:long_time}

\begin{figure}
\centering
\includegraphics[width=\linewidth]{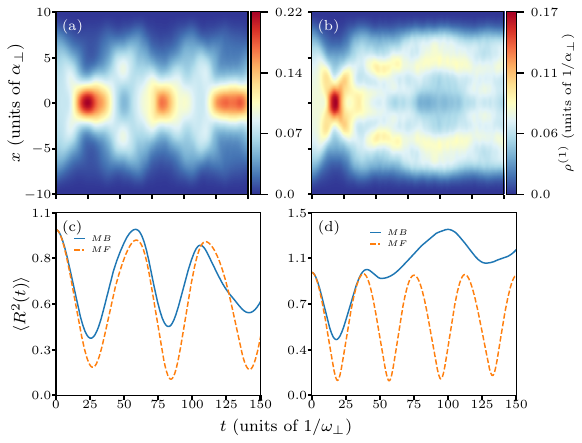}
\caption{Single-particle density evolution, $\rho^{(1)}(x,t)$, of the bosonic gas within the MB approach following an interaction quench from $g_i=0.05$ to (a) $g_f=-0.05$ and (b) $g_f=-0.1$. Besides the quantum PS, MI signatures are present for longer evolution times especially in panel (b). Time evolution of the position variance of the gas within the MF and the MB methods (see legends) for (c) $g_f=-0.05$ and (d) $g_f=-0.1$.
}
\label{fig_appendix_long-time_evol}
\end{figure}

In the main text, we discussed the correlation properties of the quantum PS for time evolutions shortly after its creation and dissolution [see, e.g., Fig.~\ref{fig1:1b-density}(b)]. 
For completeness, here the long-time dynamics of the quenched gas is demonstrated in order to visualize signatures of the subsequent MI of the background, and also further corroborate the significant correlation-induced alterations of the MB response. 
The spatiotemporal evolution of the single-particle density triggered by an interaction quench from $g_i=0.05$ to either $g_f=-0.05$ or $g_f=-0.1$ is presented in Figs.~\ref{fig_appendix_long-time_evol}(a) and ~\ref{fig_appendix_long-time_evol}(b), respectively. 
It can be seen that for relatively weak postquench attractions [Fig.~\ref{fig_appendix_long-time_evol}(a)] the quantum PS soon after its formation dissolves into a delocalized structure. Partial revivals are observed at the center around $t \sim 85$, $130$ upon reflection of the
emergent matter wave packets (arising upon disappearance of the quantum PS)
from the domain boundary.  
The long-time response is more clear at stronger postquench attractions [Fig.~\ref{fig_appendix_long-time_evol}(b)] where the quantum PS is immediately followed by an overall  spatial delocalization accompanied by dominant density excitations that manifest the MI character of the attractive background. In this case, the localization arising from the
quantum PS splinters seems to be more persistent.

It is also important to highlight the differences of the MB predictions compared to the MF ones for long evolution times. 
To do so, we utilize the position variance of the bosonic cloud defined in Eq.~(\ref{variance}). 
This is depicted in Figs.~\ref{fig_appendix_long-time_evol}(c) and ~\ref{fig_appendix_long-time_evol}(d) for the above-discussed postquench interactions. 
In the MF realm, the variance performs an oscillatory behavior of almost constant amplitude and frequency in the course of the evolution, underscoring the recurrence of the generated patterns. 
As such, it implies that within the MF approximation the gas assembles in a pattern reminiscent of a time-periodic Kuznetsov-Ma breather. 
This is in sharp contrast to the MB results, where a PS structure is generated 
and subsequent evolution may favor localization away from the centers shown in
Fig.~\ref{fig_appendix_long-time_evol}(b), rather than the above-mentioned 
MF recurrence. 
Therefore, we can infer that the inclusion of correlations not only alters the morphology of the RW excitation, but may also modify the dynamical manifestation of different RW structures. 
This is arguably another intriguing prospect for future investigations based on our results.

\section{Mean-field transition from time-periodic  breathers to the Peregrine}\label{app:transition}

In Sect.~\ref{sec:KZ_toPS_trans}, we argued that an increasing box size, $L$, with constant atom number, $N$, and quench interaction coefficients ($g_i$, $g_f$) may lead to a transition from the dynamical formation of time-periodic breathers featuring local Peregrine structures at the focusing points to a nonperiodic Peregrine waveform. 
This observation suggests that the box size can be potentially used to regulate the dynamical nucleation of the Peregrine. 
Below, we further elaborate on this crossover for varying box size and particle number along the $N$-$L$ plane. 
Here, our investigation is restricted to the MF approximation since the respective full MB simulations for larger atom numbers and/or box sizes are computationally challenging and highly time-consuming due to the increased Hilbert space dimensions. 
Moreover, the focus of the current work is to reveal the quantum properties of the PS rather than exploring the available parameter space where more complex RW-type configurations may emerge especially in the largely unexplored MB realm. 
Indeed, according to our simulations the emergence of higher-order waveforms is, in particular, favored upon considering stronger postquench attractive interactions already at the MF level (not shown for brevity).   
Nevertheless, we remark that a full investigation of the entire parameter space is desirable in future MB studies.

\begin{figure}[t!]
\centering
\includegraphics[width=\linewidth]{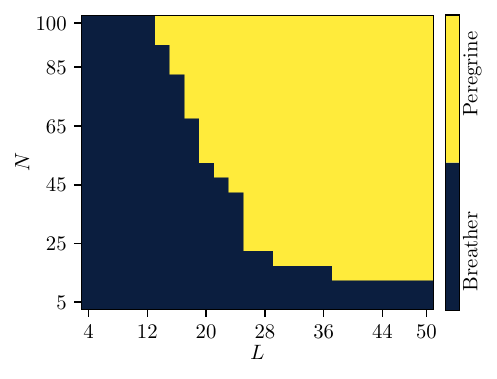}
\caption{Phase diagram depicting the parameter regions across the $N$-$L$ plane where either time-periodic breather structures (blue shaded area) or Peregrine waveforms (yellow area) occur  after an interaction quench from $g_i=0.05$ to $g_f=-0.05$. 
It is evident that for fixed $N$, Peregrine nucleation is favored for increasing box size. 
In all cases, the system is confined in a box potential and it is prepared in its ground state with $g_i=0.05$.}
\label{fig-appendix-diagram}
\end{figure}

The ensuing diagram demonstrating the parameter regions where either time-periodic configurations or a PS waveform  occur in the $N$-$L$ plane is presented in Fig.~\ref{fig-appendix-diagram}. 
Overall, it can be seen that larger box sizes for fixed $N$ facilitate  the nucleation of the PS; see the yellow shaded area in Fig.~\ref{fig-appendix-diagram}. 
The only exception is for $N \leq 8$, where we solely observe time-periodic breathers at least for the considered box sizes. 
On the other hand, it turns out that for $L<14$ only time-periodic configurations emerge, while if the box size ($L>14$) is held fixed, then Peregrine generation takes place for larger $N$. 
Moreover, we remark that for either larger box sizes (with constant $N$) or increasing atom number (with fixed $L$), the emergence of the Peregrine is usually followed by a cascade of Peregrine-type waveforms located symmetrically with respect to $x=0$ (not shown).

To classify the above dynamical waveforms, we rely on different observables. 
First, we compute the time evolution of the position variance $\braket{X^2(t)}$ [Eq.~(\ref{variance})], which  captures the collective motion of the cloud. 
In the case of time-periodic structures, $\braket{X^2(t)}$ undergoes an oscillatory behavior of almost constant amplitude, otherwise this periodicity breaks, and multifrequency oscillations occur with varying amplitude. 
In addition, having identified a nonperiodic trend of $\braket{X^2(t)}$ we inspect the density distribution and evaluate whether there is a central localized and high-amplitude structure with side nodes at the first minimum of the underlying variance. 
For our simulations, this structure possesses at least four times larger amplitude compared to its background. 
Next, we extract the phase of the wave function at the time instant of formation of the aforementioned localized configuration and ensure that there is an approximately $\pi$ phase jump between the core of the structure and its side nodes. 
The above criteria enable us to infer that the underlying waveform is of Peregrine type. 

\begin{figure*}
\centering
\includegraphics[width=\linewidth]{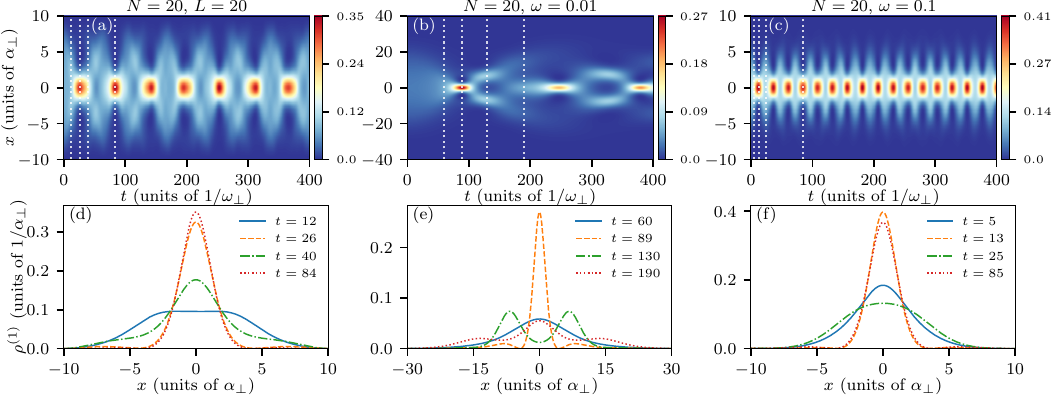}
\caption{(a)-(c) Density dynamics in the MF approximation of $N=20$ bosons subjected to an interaction quench from $g_i=0.05$ to $g_f=-0.05$. 
The system is confined in a (a), (d) box potential of length $L=20$, and in a harmonic trap of frequency (b), (e) $\omega=0.01$ and (c), (f) $\omega=0.1$. 
(d)-(f) Density profiles at selected time instants (see legends) of the dynamics depicted in panels (a)-(c), respectively. The dotted vertical white lines mark the aforementioned time-instants.  
A time-periodic breather reminiscent of the Kuznetsov-Ma appears in the presence of a box potential and a harmonic trap of frequency $\omega=0.1$, while a local Peregrine configuration emerges in the case of $\omega=0.01$. 
The system is initiated in its ground state with $g_i=0.05$.}
\label{fig-appendix-trap}
\end{figure*}

\section{Impact of a harmonic trap}\label{app:harmmonic_trap}

In the main text, we discussed the dynamical nucleation of the PS configuration stemming from the combination of the quench dynamics to attractive interactions of the bosonic sample and the finite size box confinement. 
It is indeed well-established that the PS, as a doubly-localized structure, appears in the case of attractive interactions~\cite{peregrine_water_1983}. 
However, the role of the external trap which practically shapes the initial condition is worthwhile to additionally
consider. 
While this topic has been partially considered
before at the MF level~\cite{Charalampidis2018Rogue},
we revisit the subject leveraging also additional
computations within the MF setup considered herein.
Accordingly, below, we elucidate the role of a harmonic trap in the emergent non-equilibrium dynamics of the bosonic system employed in the main text. 
This investigation aims to expose the impact of the initial condition curvature to the emergent dynamics. 
Specifically, we first obtain the underlying ground state configuration of $N=20$ atoms with $g_i=0.05$ and trigger the evolution by quenching the interaction strength to $g_f=-0.05$. 
As expected, the ground state is a Gaussian configuration (and not a Thomas-Fermi due to the relatively small atom number) which is practically shaped by the presence of the harmonic trap. 
Notice here the difference with the ground state in the box potential, see also Fig.~\ref{fig3:density profile}(a), which corresponds to a nearly homogeneous distribution with smooth edges at the box boundaries.

Figure~\ref{fig-appendix-trap} presents the density evolution within the MF approximation following the interaction quench to $g_f=-0.05$ for $N=20$ atoms confined in the box of length $L=20$ [Fig.~\ref{fig-appendix-trap}(a)] and a harmonic trap of frequency $\omega=0.01$ [Fig.~\ref{fig-appendix-trap}(b)] and $\omega=0.1$ [Fig.~\ref{fig-appendix-trap}(c)]. 
As discussed in the main text, a time-periodic breather of almost constant amplitude and frequency reminiscent of the Kuznetsov-Ma configuration is observed within this parameter regime in the case of the box potential [Fig.~\ref{fig-appendix-trap}(a)]. 
The distributions at the contraction points of the cloud are spatially localized with a large amplitude compared to their background, while signatures of side nodes are evident [Fig.~\ref{fig-appendix-trap}(d)]. 
An almost $\pi$ phase jump between these side nodes and the core takes place further confirming that at the contraction points the ensuing waveforms have PS characteristics.

However, we find that the dynamical response in the harmonic potential depends strongly on the confinement strength. 
In particular, for weak traps such as $\omega=0.01$ [Fig.~\ref{fig-appendix-trap}(b)] the focusing effect due to the quench produces a doubly localized Peregrine-like structure around $t \approx 90$, which dissolves into two- and three-hump structures; 
see also the density profiles in Fig.~\ref{fig-appendix-trap}(e). 
Notice the nicely resolved PS configuration whose tails are curved due to the presence of the external trap~\cite{Romero_theory}. 
It also features a $\pi$ phase jump between its core and its side nodes (not shown). 
Furthermore, at longer evolution times, recurrences of the above-described dynamics occur. 
On the other hand, for tighter traps the quench yields a time-periodic spatially localized configuration around the trap center at the contraction points of the cloud [Fig.~\ref{fig-appendix-trap}(c)]. This response resembles again the Kuznetsov-Ma breather with local high-amplitude Peregrine-type structures when the cloud contracts. 
Here, also the phase jump between the core and the side nodes of these temporally localized structures is a multiple of $\pi$. 
The overall cloud features contraction and expansion dynamics with an oscillation frequency $\omega_{{\rm osc}} \approx 2\omega$, which manifests the excitation of a collective breathing mode. 
Summarizing, we can infer that the presence of the trap does not preclude the emergence of RW configurations (at least in the considered parametric regime). 
Notably, the trap strength allows to transition from time-periodic breathers to doubly localized, local Peregrine-like structures.

\section{Convergence of the MB simulations}\label{app:converge}

To address the dynamical generation of the quantum PS, we numerically solve the 
time-dependent Schr\"odinger equation of the considered single-component bosonic 
gas utilizing the MCTDHB method~\cite{mctdhb-alon-PhysRevA.77.033613,alon2007unified}, relying on the generalized ML-MCTDHX~\cite{cao2017unified,mlx_jcp_2013,njp_mlx_2013} computational package. 
MCTDHB is a flexible variational approach for calculating the stationary and, most importantly, the nonequilibrium quantum dynamics of ultracold bosonic settings.
It achieves this by expressing the total MB wave function in a time-dependent and variationally
optimized basis [see also Sec.~\ref{Sec:MB_approach} and Eq.~(\ref{psi_ansatz_eq})].  
This crucial facet allows to capture the interparticle correlations of the system under study by exploiting a computationally feasible basis size. 
The basis efficiently spans the relevant subspace of the Hilbert space at each time instant; see also the reviews of Refs.~\cite{Lode_review,mistakidis2023few} for further details and applications. 

The underlying Hilbert space truncation is mainly dictated by the number of used SPFs, $M$. 
Moreover, within our implementation we rely on a sine discrete variable representation 
as a primitive basis, with $\mathcal{M}=250$ grid points to represent the spatial part of the SPFs. 
This DVR intrinsically introduces hard-wall boundary conditions at both edges of the numerical grid, imposed herein at $x= \pm10$. 
The ground states of our MB system are obtained through the imaginary time propagation method within MCTDHB. The emergent nonequilibrium quantum dynamics is simulated by propagating the MB wave function of Eq.~(\ref{psi_ansatz_eq}) in real time, subjected to the Hamiltonian of Eq.~(\ref{hamiltonian_eq}) within the MCTDHB equations of motion.

\begin{figure}
\centering
\includegraphics[width=\linewidth]{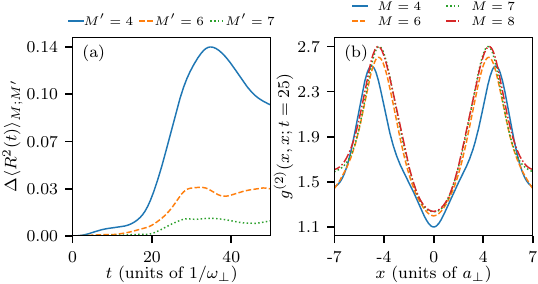}
\caption{(a) Dynamics of the relative error of the position variance [Eq.~(\ref{variance_deviation})] and (b) profiles of the diagonal of the two-body coherence [Eq.~(\ref{coh2})] at the time instance of the PS formation for different numbers of considered orbitals (see legends). 
Convergence of both one- and two-body observables takes place by increasing the number of orbitals.
}
\label{fig:convergence_check}
\end{figure}

To guarantee the convergence of our MB simulations, we ensure that all observables of interest become to a certain degree insensitive upon increasing the number of the employed orbitals ($M$).
Below, we elaborate on the convergence behavior of the position variance in the course of the evolution and the two-body coherence function at the time of the quantum PS formation for the principal considered system.  
The latter consists of $N=20$ bosons confined in a box of length $L=20$ following an interaction quench from $g_i=0.05$ to $g_f=-0.05$. 
The same convergence criteria have been imposed for the other protocols, but are omitted here for brevity.  

More concretely, we analyze the absolute deviation of the position variance in the course of the evolution between the case of $M=8$ and other $M'$ orbital values, namely,
\begin{equation}
\Delta \braket{R^2(t)}_{M;M'} = \frac{\abs{\braket{R^2(t)}_{M} - \braket{R^2(t)}_{M'}}} {\braket{R^2(t)}_{M}},\label{variance_deviation}   
\end{equation}
where $\braket{R^2(t)}_M=\braket{X^2(t)}_{M}/\braket{X^2(0)}_{M}$. 
The time evolution of $\Delta \braket{R^2(t)}_{M;M'}$ is depicted in Fig.~\ref{fig:convergence_check}(a) for different number of orbitals, $M'$. 
By inspecting $\Delta \braket{R^2(t)}_{M;M'}$, a systematic convergence trend is observed for larger $M'$.  For instance, $\Delta \braket{R^2(t)}_{8;M'}$ for $M'=4$ ($M'=7$) attains a maximum of the order of $14 \%$ [$\leq 1 \%$], exemplifying the numerical convergence of the simulations at the one-body level. 
Next, we study the behavior of the diagonal of the second-order coherence function $g^{(2)} (x, x;t=25)$ at the time instant of the Peregrine formation for varying orbital number [see Fig.~\ref{fig:convergence_check}(b)].  
This is arguably a more difficult measure to explore convergence as compared to $\Delta \braket{R^2(t)}_{M;M'}$, since it is a two-body observable. 
The profile of $g^{(2)} (x, x;t=25)$ evinces a clear strong bunching behavior [$g^{(2)} (x=\pm 4, x \pm 4;t=25) >1$] at each side of the quantum PS, while a progressive convergence trend occurs for increasing $M'$. 
Therefore, we demonstrated the numerical convergence of our MB simulations at both the one- and two-body levels.

\bibliography{literature.bib}

\end{document}